\documentclass[
 reprint,
superscriptaddress,
 amsmath,amssymb,
 aps,
superscriptaddress,
 amsmath,amssymb,
 aps,longbibliography
]{revtex4-1}

\usepackage{graphicx}
\usepackage{dcolumn}
\usepackage{bm}
\usepackage{amsmath,amssymb,amstext,amsthm}
\usepackage{braket}
\usepackage{bbold}
\usepackage{bbm}
\usepackage{xcolor}
\usepackage{ gensymb }
\usepackage{soul}
\setstcolor{red}
\usepackage{ mathrsfs }
\usepackage{cancel}
\usepackage{multirow}

\usepackage{hyperref}

\hypersetup{colorlinks,linkcolor=blue,citecolor=blue,urlcolor=blue}

\begin{document}

\preprint{APS/123-QED}
\setlength{\abovedisplayskip}{1pt}
\title{Quantitative diagnosis of amyloid without Congo red staining using polarized light microscopy}



\author{Owen Lailey} 
\email{oalailey@uwaterloo.ca}
\affiliation{Institute for Quantum Computing, University of Waterloo,  Waterloo, ON, Canada, N2L3G1}
\affiliation{Department of Physics and Astronomy, University of Waterloo, Waterloo, ON, Canada, N2L3G1}

\author{Maria Agustina Alais}
\affiliation{Institute for Quantum Computing, University of Waterloo,  Waterloo, ON, Canada, N2L3G1}
\affiliation{Department of Laboratory Medicine and Molecular Diagnostics, Sunnybrook Health Sciences Centre, Toronto, ON, Canada}

\author{Liuhe Wang} 
\affiliation{Institute for Quantum Computing, University of Waterloo,  Waterloo, ON, Canada, N2L3G1}
\affiliation{Department of Physics and Astronomy, University of Waterloo, Waterloo, ON, Canada, N2L3G1}

\author{Pinki Chahal}
\affiliation{Department of Physics, University at Buffalo, State University of New York, Buffalo, New York 14260, USA}

\author{David G. Cory}
\affiliation{Institute for Quantum Computing, University of Waterloo,  Waterloo, ON, Canada, N2L3G1}
\affiliation{Department of Chemistry, University of Waterloo, Waterloo, ON, Canada, N2L3G1}

\author{Timothy Khoo} 
\affiliation{Department of Physics and Astronomy, University of Waterloo, Waterloo, ON, Canada, N2L3G1}

\author{Ekaterina Olkhov-Mitsel}
\affiliation{Department of Laboratory Medicine and Molecular Diagnostics, Sunnybrook Health Sciences Centre, Toronto, ON, Canada}

\author{Dusan Sarenac}
\affiliation{Department of Physics, University at Buffalo, State University of New York, Buffalo, New York 14260, USA}

\author{Dmitry A. Pushin}
\email{dmitry.pushin@uwaterloo.ca}
\affiliation{Institute for Quantum Computing, University of Waterloo,  Waterloo, ON, Canada, N2L3G1}
\affiliation{Department of Physics and Astronomy, University of Waterloo, Waterloo, ON, Canada, N2L3G1}

\author{Jelena Mirkovic}
\email{jelena.mirkovic@sunnybrook.ca}
\affiliation{Department of Laboratory Medicine and Molecular Diagnostics, Sunnybrook Health Sciences Centre, Toronto, ON, Canada}
\affiliation{Department of Chemistry, University of Waterloo, Waterloo, ON, Canada, N2L3G1}
\affiliation{Department of Laboratory Medicine and Pathobiology, University of Toronto, Toronto, ON, Canada}

\date{\today}


\pacs{Valid PACS appear here}


\begin{abstract}
Amyloidosis is a protein misfolding disease caused by the deposition of large, insoluble aggregates (amyloid fibrils) of protein in a tissue, which has been associated with various conditions, such as lymphoid disorders, Alzheimer's disease, diabetes mellitus type 2, chronic inflammatory processes, and cancers. Amyloid fibrils are commonly diagnosed by qualitative observation of green birefringence from Congo red stained biopsy tissue samples under polarized light, a technique that is limited by lack of specificity, dependence on subjective interpretation, and technical constraints.  Studies emphasize the utility of quantitative polarized light microscopy (PLM) methodology to diagnose amyloid fibrils in Congo red stained tissues. However, while Congo red enhances the intrinsic birefringence of amyloid fibrillar structures, there are significant disadvantages such as the appearance of multiple non-green colors under polarized light and binding to other structures, which may result in misdiagnoses with Congo red dye and inconclusive explanations. In this work, we present an improved PLM methodology for quantitative detection of amyloid fibrils without requiring Congo red staining. We perform PLM measurements on four tissues: abdominal subcutaneous tissue biopsy, duodenal biopsy, thyroid biopsy, and breast biopsy, both with Congo red stain and H\&E stain, and through Fourier analysis quantify birefringence, birefringent axis orientation, dichroism, optical activity, and relative amyloid density. These results emphasize a quantitative analysis for amyloid diagnosis rooted in Fourier signal harmonics that does not require Congo red dye and paves the way for rapid, simple, and accurate diagnosis of amyloid fibrils.


\end{abstract}
\maketitle

\section{Introduction}
 Amyloids, historically named for their ‘starch-like’ appearance upon iodine staining, are aggregates of misfolded, self-assembled proteins that form highly insoluble structures known as amyloid fibrils. These fibrils exhibit a characteristic cross-$\beta$ fiber diffraction pattern under X-ray. Pathogenic fibril deposition disrupts normal tissue architecture, leading to organ dysfunction and a series of disorders collectively known as amyloidosis~\cite{eisenberg2017structural, bakeramyloidoses, desikan1997incidence, patel2015clinicopathological, schonland2012immunohistochemistry}. Of clinical significance, a certain kind of amyloid fibrils, amyloid-$\beta$, are a defining neuropathological feature of Alzheimer’s disease. Additionally, amyloid formation has been implicated to be related with cancer  \cite{mizejewski2017breast,sengupta2023p53,iwahashi2022cytoplasmic,ghosh2017p53}, lymphoid disorders \cite{sanchorawala2006amyloidosis,dhakal2015localized,fujiwara2009primary}, diabetes mellitus type 2 \cite{marzban2003islet,abedini2013mechanisms,hoppener2000islet}, and chronic inflammatory conditions \cite{scarpioni2016secondary,cunnane2001amyloid}. Nonetheless,  it is important to note that amyloid formation is not just associated with disease states. It is a well-defined structural form of the protein, alternative to the native state, that may be adopted by many polypeptide sequences~\cite{fowler2007functional, greenwald2010biology}.  Certain amyloids may play functional roles in the human body, such as in storage of peptide hormones and necroptosis of cells \cite{brown2021functional}. 
 
It has been indicated that the diagnosis of amyloidosis is often delayed due to the the nondescript main symptoms of such diseases~\cite{lin2015patient,hayman2001primary,gertz1989primary}.
Timely diagnosis and treatment plays a crucial role in the management of amyloidosis~\cite{rapezzi2009systemic, merlini2003molecular,rh1997systemic, gertz1999amyloidosis} in order to prevent severe tissue damage and dysfunction~\cite{maurer2017addressing,wechalekar2016systemic,dietrich2010treatment}.
However, with current methods, patients commonly need to see multiple physicians, in specific referral centers with well-trained pathologists, increasing the time to diagnosis~\cite{maurer2017addressing}. 


Amyloids can be seen on standard hematoxylin and eosin (H\&E) stain, appearing as amorphous eosinophilic deposits.  However, H\&E staining alone is not specific enough to differentiate from visually similar tissue changes, such as hyaline deposits. Therefore, other more specific stains are required to confirm the diagnosis. Despite the development of more specific immunohistochemical stains for certain amyloids, such as amyloid-$\beta$ associated with Alzheimer's disease~\cite{hardy2002amyloid,gremer2017fibril,westermark2002amyloid,ow2014brief, korczyn2024alzheimer, mizejewski2017breast}, the gold standard for the detection of amyloid fibrils in most histopathology laboratories is still to stain tissue samples with Congo red dye
~\cite{yakupova2019congo}.   Congo red stain enhances the intrinsic birefringence of amyloid fibrils~\cite{gremer2017fibril, eisenberg2017structural, taylor1974determination, wolman1965cause, smith2012giant,csen2012digitally} and when observed under a microscope with crossed polarizers ($90\degree$~offset), Congo red stained amyloids exhibit `apple green' birefringence, allowing for trained pathologists to qualitatively identify amyloid fibrils~\cite{jin2003imaging,cohen1986general, elghetany1988methods, rosenblum2002structure}. Congo red staining with polarized microscopy is limited by lack of specificity for amyloid deposits, variable sensitivity, dependence on subjective interpretation, and technical constraints.~\cite{el2019improving,howie2009optical}. In standard clinics, the lenses used in the microscopes are manufactured with strain, which will affect the observations under polarized light and leads to observations of several colours~\cite{el2019improving, howie2008physical, howie2009optical, howie2019origins}. This can result in misdiagnosis of amyloid fibrils and confusion among scientists that green and only green is indicative of amyloid when stained with Congo red~\cite{bowen2012amyloidosis,senecal2023amyloidosis}. Additionally, `apple green' birefringence is a qualitative description, dependent on the observers' subjective judgment, in need of replacement with quantitative methods as motivated by work strongly correlating amyloid density, in certain cases, with disease outcomes~\cite{sengupta2023p53, iwahashi2022cytoplasmic, ghosh2017p53, schonland2012immunohistochemistry, patel2015clinicopathological, mizejewski2017breast, rogers2005electric,rogers2005measuring,clement2014evaluation,beach2012striatal}. Furthermore, there is ample evidence that Congo red binds to several other substances other than amyloid, confirming that detecting birefringence with Congo red stained samples is not specific for amyloid ~\cite{yakupova2019congo,bayer2002amyloidosis,horobin1970staining,horobin1980structure,lendrum1972renal,westermark19991}. 



\begin{figure}
    \centering\includegraphics[width=1\linewidth]{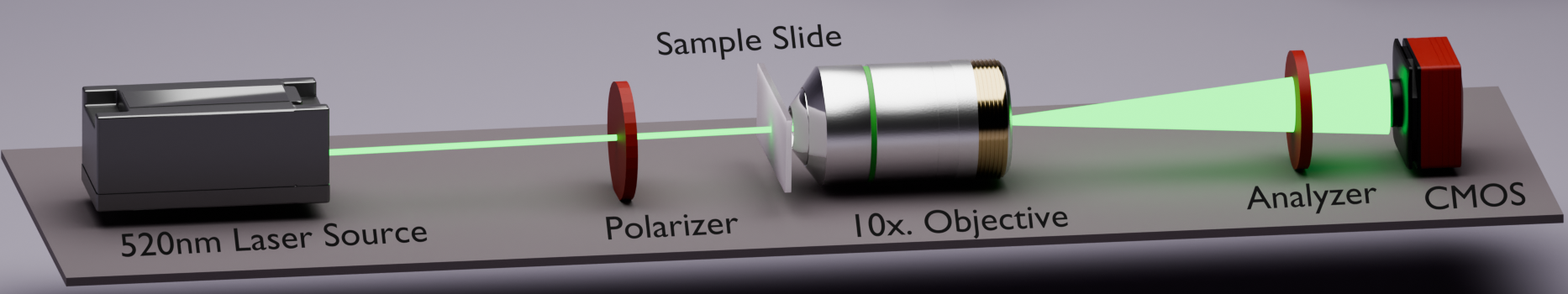}
    \caption{A simplified schematic of the PLM optical setup. A 520 nm green laser passes through a linear polarizer on a rotation mount before interacting with the sample.  A 10x objective, that is strain- and stress-free, collects light after the sample, 
    before the other linear polarizer on a rotation mount (the analyzer). Lastly, images are acquired using a CMOS camera with proper tube lens connected to a laptop. 
    }
    \label{fig:schem}
\end{figure}


In this work we have implemented a polarized light microscopy (PLM) system capable of detecting and quantifying birefringent material such as amyloid in biological samples without the need for Congo red staining. We apply PLM techniques~\cite{he2021polarisation, wolman1975polarized, jin2003imaging, tuchin2016polarized, changoor2011polarized,westreich2019novel}  to quantify birefringence, birefringent axis orientation, dichroism, optical activity, and relative amyloid density through Fourier analysis. We measure birefringence contrast maps using a polarizer/analyzer pair for standard Hematoxylin and Eosin (H\&E) stained biological tissue samples and obtain results comparable to Congo red stained samples with the expected reduced contrast~\cite{wolman1965cause, taylor1974determination, howie2009optical, howie2019origins}. This work opens the door for techniques identifying and determining amyloid density based on its intrinsic birefringence and complements other methods, such as those using deep learning~\cite{yang2024virtual}, to provide alternatives to Congo red staining. 

\section{Materials and Methods}
\label{meth}
\subsection{Human tissue samples}
Samples were obtained from Sunnybrook Hospital (Toronto, Canada) with appropriate institutional ethics approval. Hospital pathology files were searched for term ``amyloid" to retrospectively identify cases of amyloidosis. Four select cases of amyloidosis to include several different tissue sites and at least two different amyloid subtypes were included : 1) duodenal biopsy from patient with multiple myeloma, 2) abdominal subcutaneous tissue biopsy from a patient with macroglossia and elevated serum kappa light chains, 3) thyroid biopsy from patient with medullary thyroid carcinoma, and 4) breast biopsy from patient with mammary amyloidosis with light chain immunoglobulin type amyloid (AL amyloid) detected by mass spectrometry.  Two consecutive $4~\mu$m thick sections are prepared from retrieved formalin fixed paraffin embedded (FFPE) tissue blocks, mounted on glass slides, and stained with standard H\&E and Congo Red stains, respectively~\cite{feldman2014tissue} and ~\cite{yakupova2019congo}. 
Two pathologists (JM, MAA) identified several regions of interest (ROIs) (either amyloid or normal tissue) on Congo red stained slides and the corresponding areas on the H\&E stained slides using a Nikon Eclipse E400 Biological Microscope. These samples were then placed within our PLM setup described below.  

\begin{figure*}
    \centering\includegraphics[width=1\linewidth]{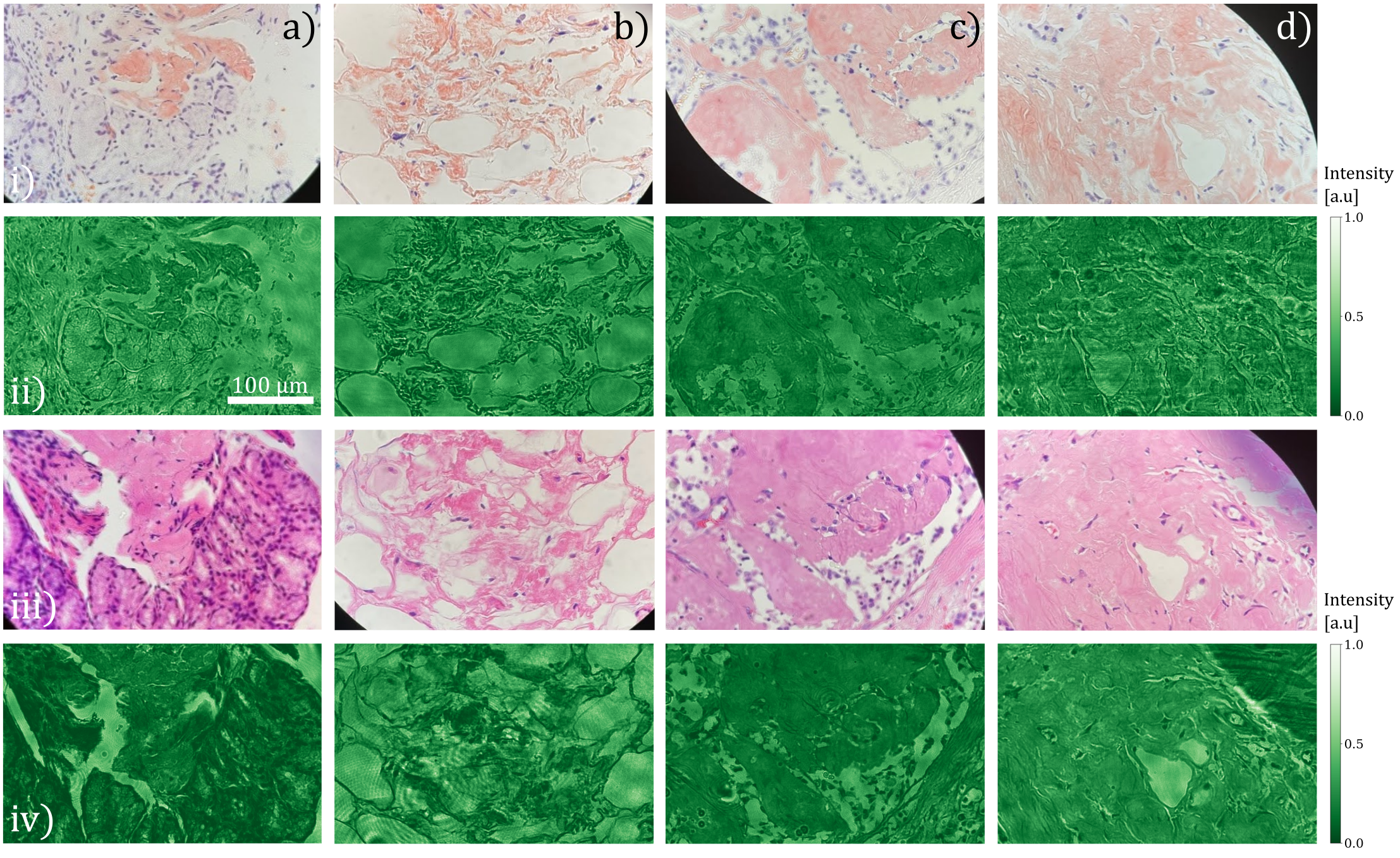}
    \caption{Conventional light microscope images at 40x magnification (i and iii) alongside the corresponding image taken with our custom PLM system in the parallel polarizer configuration (ii and iv). Columns a-d) represent four different samples: duodenal biopsy, abdominal subcutaneous tissue biopsy, thyroid biopsy, and breast biopsy, with eosinophilic amorphous material confirmed to be amyloid. Rows i-ii) are for Congo red stained samples and iii-iv) are for H\&E stained samples.}
    \label{fig:micro}
\end{figure*}

\subsection{Polarized light microscopy system}
We implemented a custom PLM system as shown in the schematic in Fig.~\ref{fig:schem}. A 
Fabry-Perot benchtop 520~nm (green light) laser source 
is operated at 5~mW and fiber coupled to a collimator. 
A fixed linear film polarizer and a 520~nm zero-order quarter-wave plate (not shown in Fig.~\ref{fig:schem}) are placed before the rotating linear polarizer, which is mounted on a rotation stage driven by a motor controller. The sample slide is mounted to a triple axis 
linear translation stage for sample positioning and focusing. 
After the sample, a 10x stress- and strain-free objective collects the transmitted light, followed by another linear polarizer (the `analyzer') within a continuous rotation mount (the same as for the polarizer) operated by a motor controller. A 
CMOS camera 
with tube lens is used for all image acquisition. The entire setup is in an enclosure composed of black anodized aluminum and blackout nylon fabric with
polyurethane coating to reduce background light, on top of a 
optical table with active isolation to reduce vibrations. 

All image acquisition is automated with custom Python scripts. The script begins when the pathologist has focused the image 
and sets the exposure time. The polarizer/analyzer pair are initially aligned (parallel position) and rotated together 360\degree~in 10\degree~steps, acquiring 37 images. For each image acquisition, multiple-exposure High Dynamic Range (HDR) imaging techniques are employed to increase the dynamic range of the images and reduce the number of over/under-exposed pixels~\cite{reinhard2020high}. Afterwards, this measurement is repeated with the analyzer rotated 90\degree~with respect to the polarizer (crossed position). Finally, 37 more images are acquired for a fixed polarizer orientation while the analyzer rotates 360\degree, used for quantifying optical activity. Each image has a resolution of 1216 by 1936 pixels, corresponding to an image size of about $230~\mu$m by $370~\mu$m, where a pixel size of about $0.2~\mu$m is obtained via calibration with standard Thorlabs USAF resolution targets. 

\subsection{Analysis methods}

\begin{figure*}
    \centering\includegraphics[width=1\linewidth]{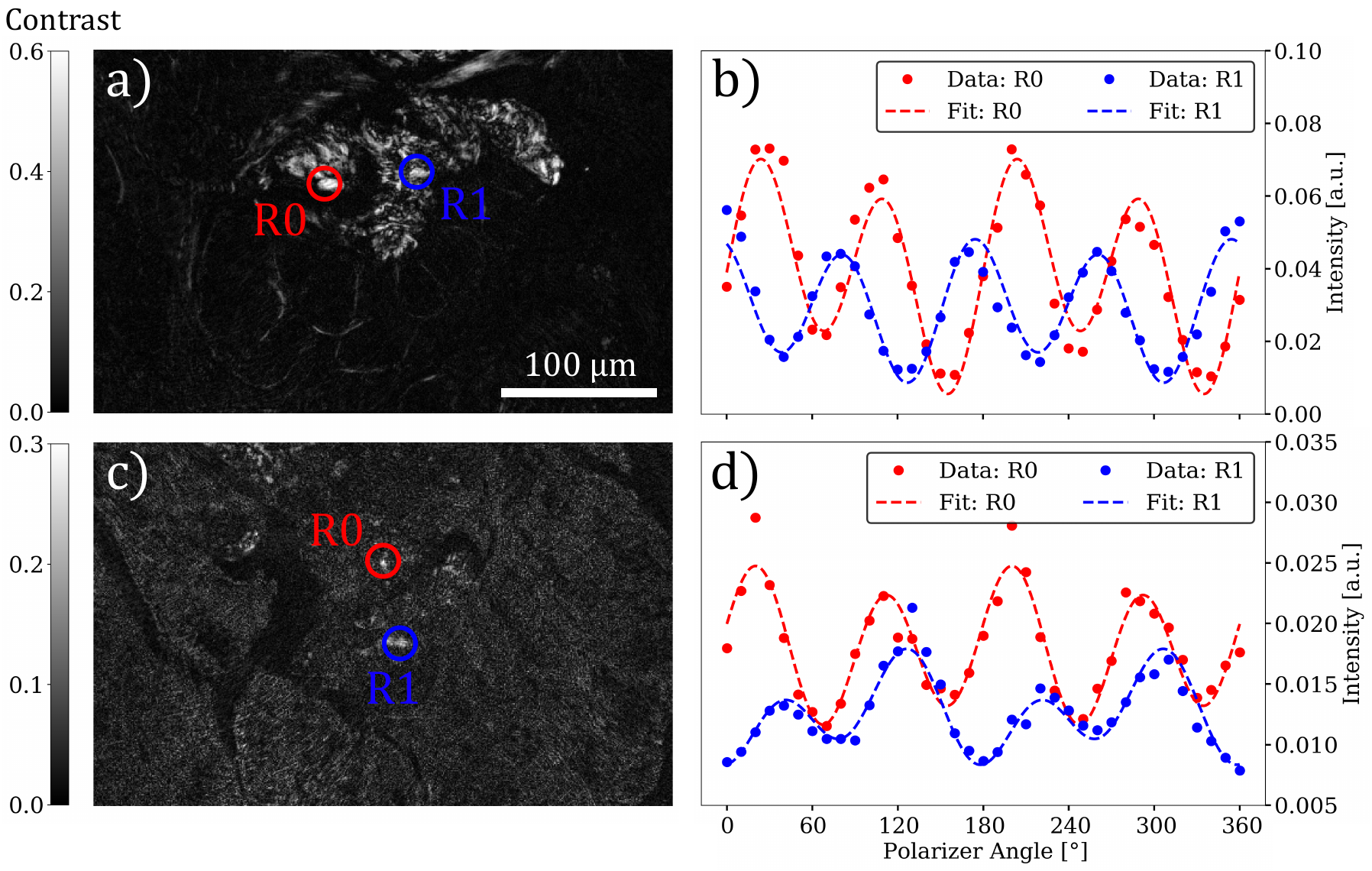}
    \caption{a) The computed birefringence contrast for the Congo red stained duodenal tissue sample. Labels R0 and R1 correspond to analyzed pixels in b), plotting measured intensity T from Eq.~\ref{T} as a function of polarizer angle. The dominant $k=4$ frequency is clearly observed as expected for birefringent structures. The data is well described by Eq.~\ref{eq_fit} which considers the characteristic $k=4$ frequency of birefringence and $k=2$ frequency of dichroism. In c-d) are the analogous figures to a-b) for the H\&E stained sample. As expected, contrast is diminished but the intrinsic birefringence of amyloid fibrils is still observable.}
    \label{fig:s12_bir}
\end{figure*}

\begin{figure*}
    \centering\includegraphics[width=1\linewidth]{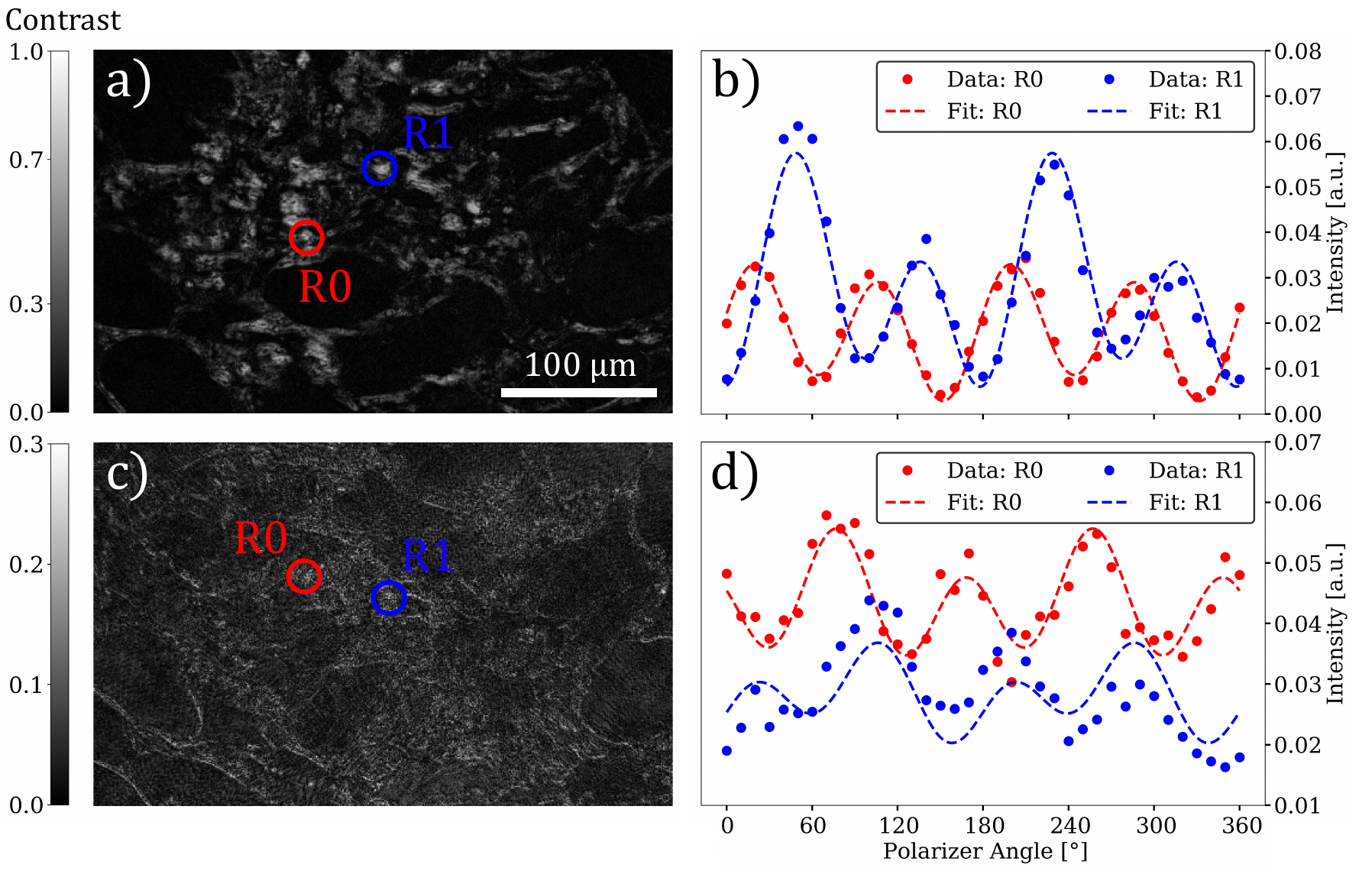}
    \caption{a) The computed birefringence contrast for the Congo red stained abdominal subcutaneous tissue sample. Labels R0 and R1 correspond to analyzed pixels in b), plotting measured intensity T from Eq.~\ref{T} as a function of polarizer angle. The dominant $k=4$ frequency is clearly observed as expected for birefringent structures. The data is well described by Eq.~\ref{eq_fit} which considers the characteristic $k=4$ frequency of birefringence and $k=2$ frequency of dichroism. In c-d) are the analogous figures to a-b) for the H\&E stained sample. As expected, contrast is diminished but the intrinsic birefringence of amyloid fibrils is still observable.}
    \label{fig:s9_bir}
\end{figure*}

\begin{figure*}
    \centering\includegraphics[width=1\linewidth]{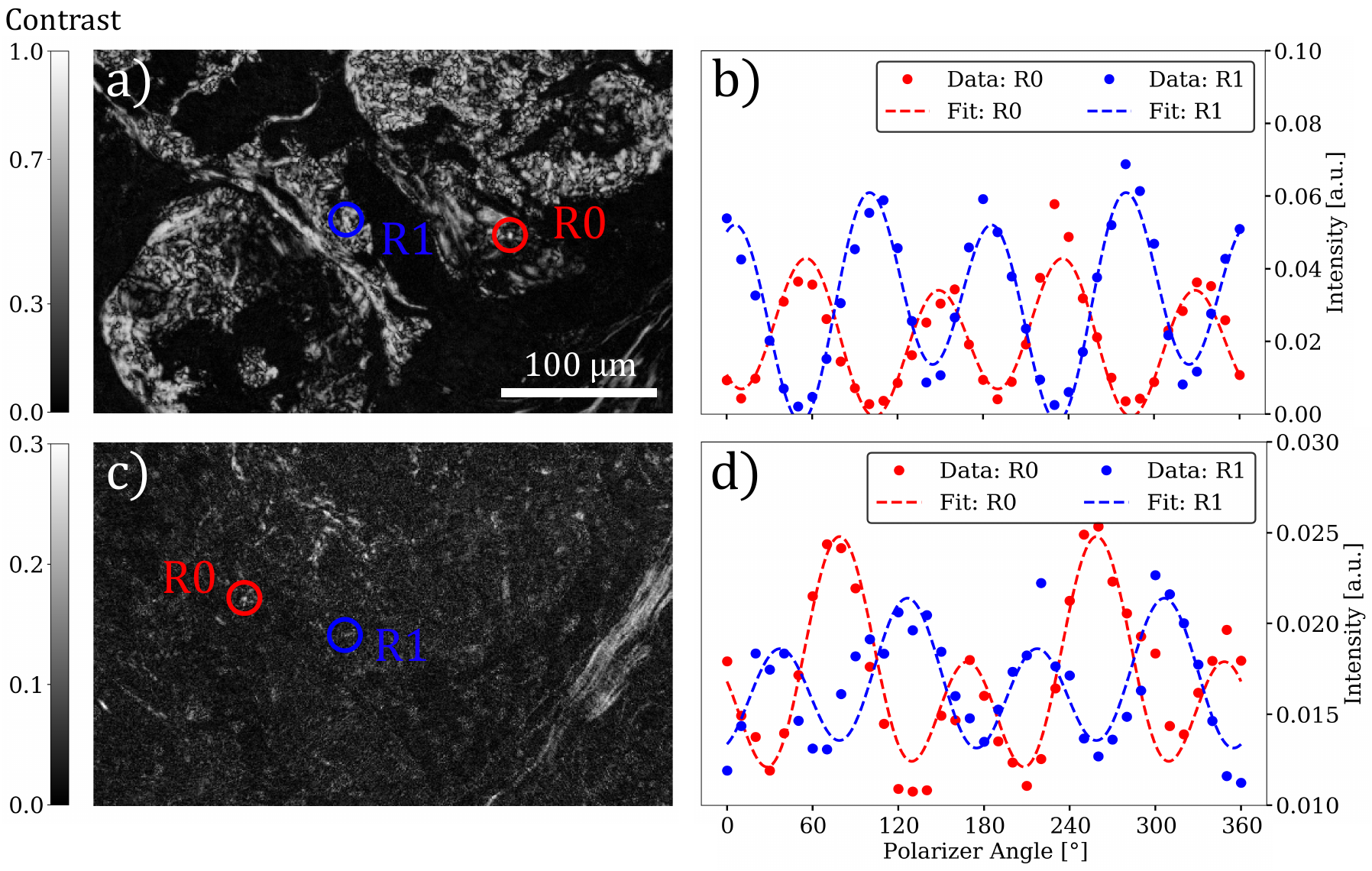}
    \caption{a) The computed birefringence contrast for the Congo red stained thyroid medullary carcinoma tissue sample. Labels R0 and R1 correspond to analyzed pixels in b), plotting measured intensity T from Eq.~\ref{T} as a function of polarizer angle. The dominant $k=4$ frequency is clearly observed as expected for birefringent structures. The data is well described by Eq.~\ref{eq_fit} which considers the characteristic $k=4$ frequency of birefringence and $k=2$ frequency of dichroism. In c-d) are the analogous figures to a-b) for the H\&E stained sample. As expected, contrast is diminished but the intrinsic birefringence of amyloid fibrils is still observable.}
    \label{fig:s10_bir}
\end{figure*}

\begin{figure*}
    \centering\includegraphics[width=1\linewidth]{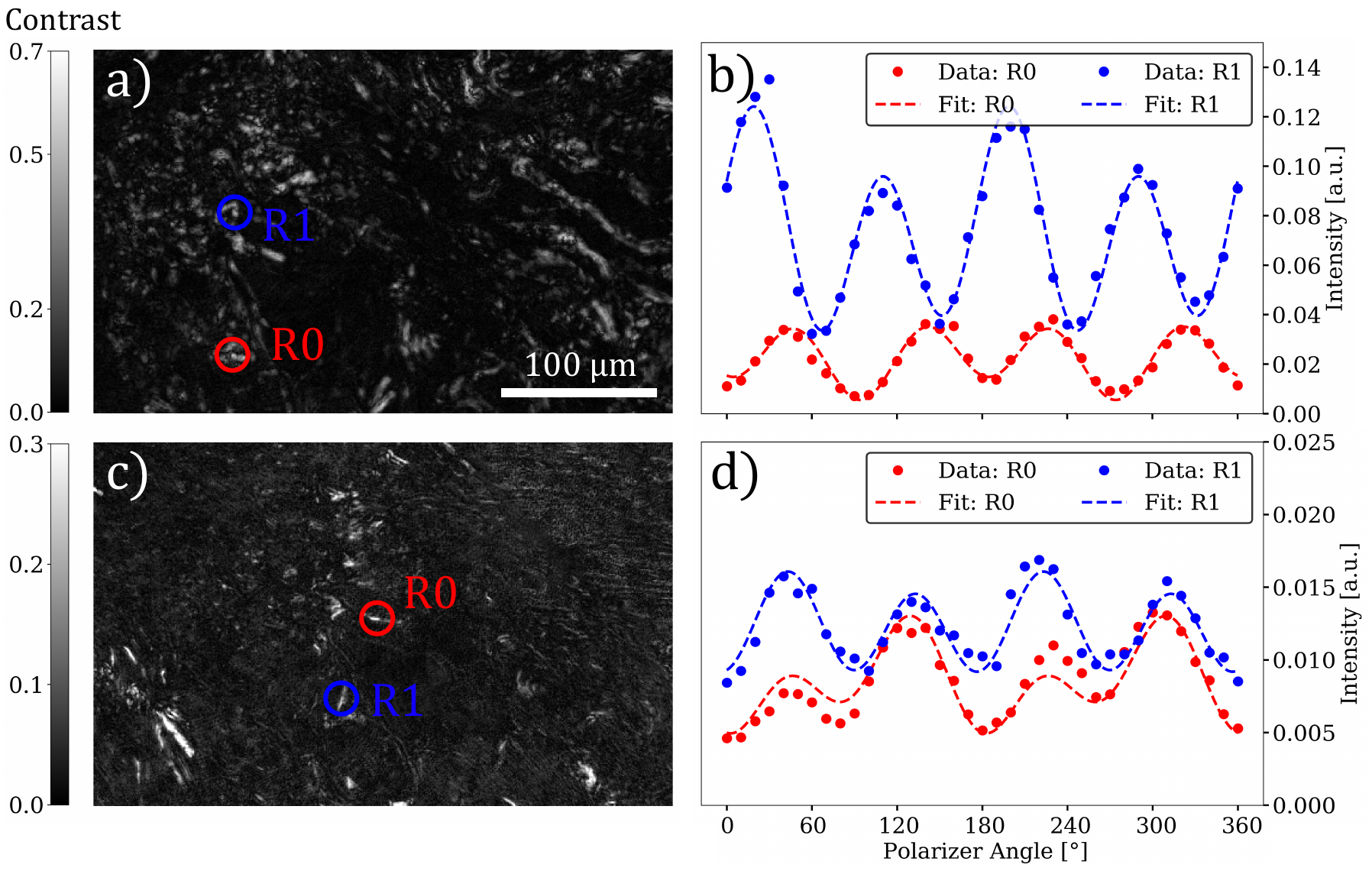}
    \caption{a) The computed birefringence contrast for the Congo red stained breast tissue sample. Labels R0 and R1 correspond to analyzed pixels in b), plotting measured intensity T from Eq.~\ref{T} as a function of polarizer angle. The dominant $k=4$ frequency is clearly observed as expected for birefringent structures. The data is well described by Eq.~\ref{eq_fit} which considers the characteristic $k=4$ frequency of birefringence and $k=2$ frequency of dichroism. In c-d) are the analogous figures to a-b) for the H\&E stained sample. As expected, contrast is diminished but the intrinsic birefringence of amyloid fibrils is still observable.}
    \label{fig:s11_bir}
\end{figure*}

Theoretically, no light will be detected at the camera when there is no sample between crossed polarizers~\cite{goldstein2017polarized}. However, a sample exhibiting birefringence located between the crossed polarizers can rotate the incoming polarized light to become elliptically polarized resulting in light detection at the camera~\cite{goldstein2017polarized}. As the crossed polarizer pair is rotated, the transmitted intensity through the sample is a Fourier sum of the various optical contributions with Fourier frequency $k = 4$ dominating for the birefringent structures:

\begin{equation}
    I \propto \sin^2{\left(\frac{k}{2}\theta\right)}\sin^2{\left(\pi t \Delta n / \lambda\right)},
    \label{deltan}
\end{equation}
where $\theta$ is the angle between the polarizer and the birefringent axis, $\Delta n$ is the difference between the fast and slow axis of the birefringent material, $t$ is the sample thickness, $\lambda$ is the wavelength of the light. and $I$ is the transmitted intensity~\cite{nelson2014interference, westreich2019novel}. Using the crossed and parallel polarizer measurements we compute the normalized transmitted crossed intensity ratio:
\begin{equation}
    T = \frac{I_{cross}}{I_{parallel} + I_{cross}},
    \label{T}
\end{equation}
where $I_{cross}$ and $I_{parallel}$ are the sets of 37 respective cross and parallel polarizer transmitted intensity images captured by the camera. We filter the images by binning 4x4 regions of pixels together and then perform the two dimensional fast Fourier transform of $T$. We compute the birefringence contrast $2B_1/A$, where $B_1$ is the amplitude of the $k = 4$ Fourier frequency component and $A$ is the mean signal. Pathologists select high contrast regions from the birefringence contrast maps to verify the $k = 4$ frequency intensity oscillations, characteristic of birefringence structure. To fit the data, we account for the signal being a Fourier sum of separate optical contributions, such as birefringence and dichroism~\cite{howie2008physical}:

\begin{equation}
\begin{split}
    T_{fit} = A & - \frac{B_1}{2}\left[1 - \cos(k_1(\theta + \phi_1))\right]\\
    & - \frac{B_2}{2}\left[1 - \cos(k_2(\theta + \phi_2))\right],
\end{split}
    \label{eq_fit}
\end{equation}
where $A$ determines the mean (background) signal, $B_1,~\phi_1,~k_1$ and $B_2,~\phi_2,~k_2$ are the respective amplitudes, phases, and frequencies of the birefringence and dichroism terms, and $\theta$ is the polarizer angle varying from 0\degree~to 360\degree. We determine $\Delta n$, the difference between the fast and slow axis of the birefringent structure, from Eq.~\ref{deltan},~\ref{T}.

As described in Ref.~\cite{jin2003imaging}, birefringence and dichroism are not distributed evenly. The dichroism contribution is described by:
\begin{equation}
    T_D =\cosh\epsilon+\sinh\epsilon \cos(k_2(\theta - \phi_2')),
    \label{eq_dic}
\end{equation}
where $T_D$ is the transmitted intensity, $\theta - \phi_2'$ is the angle between the input polarization state and the least transmitting direction of the dichroism, and $\epsilon=2\pi t \Delta k/\lambda$ represents the differential absorption with $\Delta k=2(T_{0\degree}-T_{90\degree})/(T_{0\degree}+T_{90\degree})$ defining the transmission difference between the transmitting directions. We quantify $\Delta k$ and $\phi_2'$ for all samples using Eq.~\ref{eq_fit},~\ref{eq_dic}.

From the varying brightness of the birefringent structure as the polarizer/analyzer are rotated, we define a metric to quantify the relative amount of amyloid within different regions of a single tissue sample, as well as across the various tissues themselves, which in qualitative terms spans the range `no amyloid' to `high density of amyloid'. The normal tissue (see Appendix), determines the `no amyloid' end of the metric as the average $B_1$ value for Congo red and H\&E ROI points is approximately equal to the uncertainty, demonstrating no $k=4$ Fourier frequency signal whatsoever. The `high density of amyloid' end of the metric is determined by the measured birefringent structure exhibiting the largest variation in intensity which is found in the Congo red stained breast tissue sample. To quantify these values, we compute ratios of $B_1$ amplitudes and map to a normalized scale from 0 (`no amyloid') to 1 (`high density of amyloid').

In addition to birefringence contrast maps, we analyze the birefringence orientation for determining a structure's birefringent axis, which is related to $\phi_1$ in Eq.~\ref{eq_fit}. It is expected that regions of amyloid structure will exhibit smaller variations in birefringent orientation due to their alignment whereas non birefringent structures will exhibit comparatively larger fluctuations in phase~\cite{westreich2019novel}. From Eq.~\ref{deltan}, if the highest transmission through cross polarizers occurs at angle $\theta_1$, then the structure's birefringent axis is one of $\theta_1 \pm 45\degree$. We quantify the birefringent axis for the amyloid observed in both Congo red and H\&E stained samples for all tissue samples.

After the crossed and parallel polarizer measurements, the polarizer is fixed at $0\degree$~and the analyzer is rotated $360\degree$. When the polarizer and analyzer are aligned (crossed), we expect maximum (minimum) intensity with Fourier frequency $k = 2$ as given by:
\begin{equation}
    I = C + D*\cos(k_3(\alpha + \chi)),
    \label{optical_act}
\end{equation}
where $I$ is normalized transmitted intensity at the camera, $k_3 = 2$ is the expected dominating Fourier frequency, $\alpha$ is the analyzer angle, $\chi$ is the phase shift, $C$ is the background signal, and $D$ is the amplitude. If the sample is optically active, the sample can rotate the polarization direction between the polarizer pair resulting in a nonzero $\chi$~\cite{barron2009molecular}. Furthermore, we compute the difference in refractive index for left and right circularly polarized waves from $\chi$: 
\begin{equation}
    \Delta n_{OA} = \frac{\lambda}{2\pi t}\chi,
    \label{delta_opt}
\end{equation}
where $\lambda$ is wavelength and $t$ is thickness of the sample~\cite{barron2009molecular}.

\section{Results}

\begin{figure}
    \centering\includegraphics[width=1\linewidth]{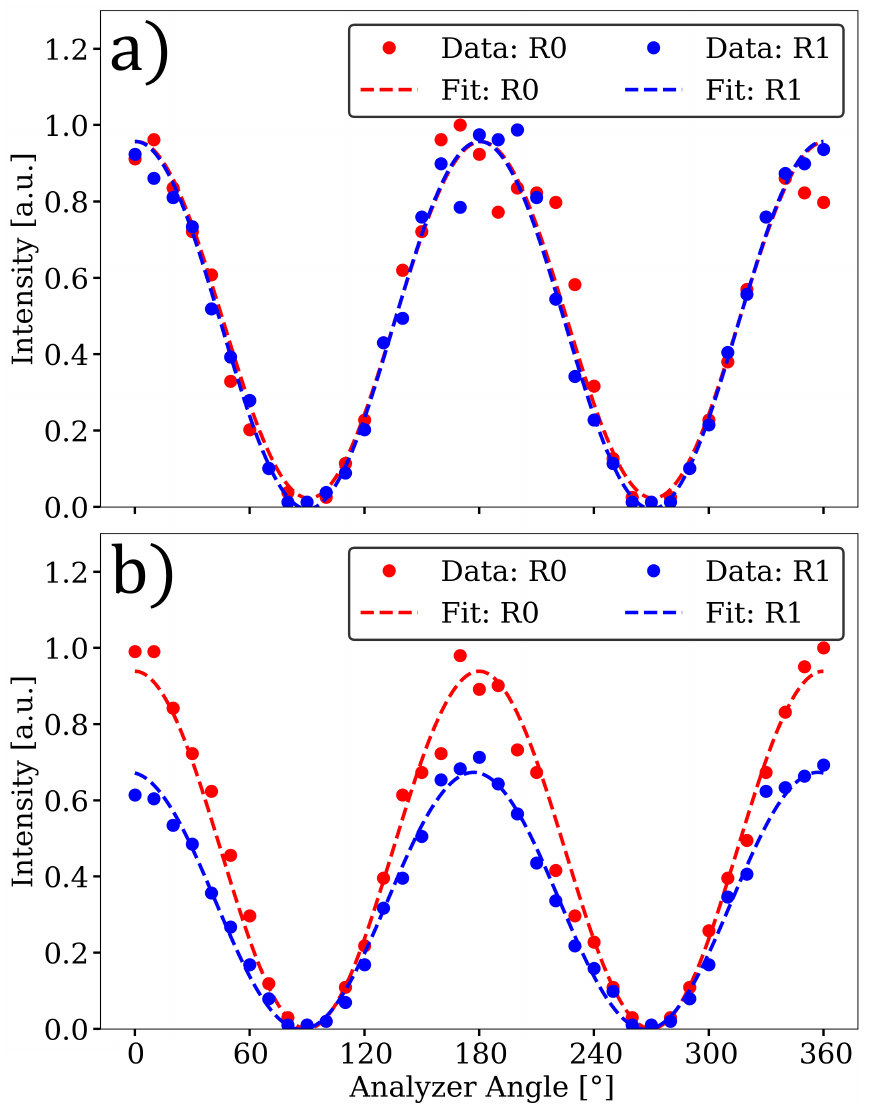}  
    \caption{a,b) Optical activity measurements for the subcutaneous cellular tissue sample whereby the polarizer is fixed and the analyzer rotates $360\degree$. Region of interest points R0 and R1 are the same as those shown in Fig.~\ref{fig:s9_bir}. The data is in excellent agreement with theory given by Eq.~\ref{optical_act}.}
    \label{fig:s9_opt}
\end{figure}

\begin{figure}
    \centering\includegraphics[width=1\linewidth]{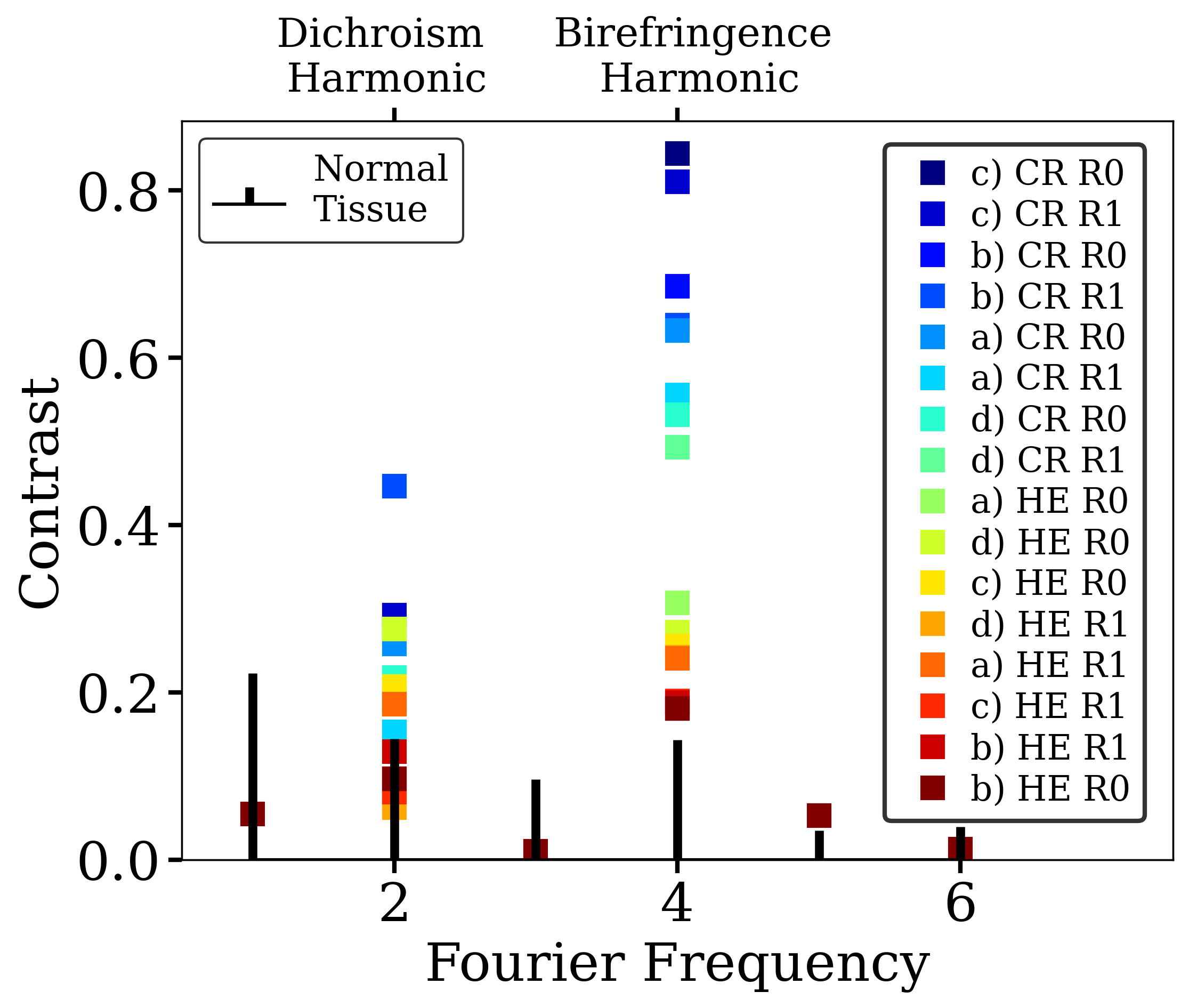}
    \caption{Contrast of the birefringence and dichroism harmonics for Congo red (CR) and H\&E (HE) stained samples a)-d) (corresponding to Fig.~\ref{fig:micro} a)-d)) for points R0, R1 from Figs.~\ref{fig:s12_bir},~\ref{fig:s9_bir},~\ref{fig:s10_bir},~\ref{fig:s11_bir}. The reference Fourier spectrum for a normal tissue region (Congo red R1 from Fig.~\ref{fig:s12_normal} in the Appendix) is shown in solid black, indicating a noisy Fourier signal and lacking the dominate birefringence harmonic. We also plot in dark red the Fourier spectrum in this range for the lowest measured birefringence contrast ROI (H\&E R0 from Fig.~\ref{fig:s9_bir}) indicating dominating birefringence and dichroism harmonics in comparison to the $k=1,~3$ contributions, which are large in the normal tissue.}
    \label{fig:freqs}
\end{figure}

Shown in Fig.~\ref{fig:micro} are the conventional light microscope images for all tissue samples described in section~\ref{meth}, alongside parallel polarizer images captured with the PLM system. The duodenal tissue sample is shown in Fig.~\ref{fig:micro}a i)-iv) with the region shown indicating an ROI from the pathologist to contain amyloid and adjacent normal tissue.  H\&E image (Fig.~\ref{fig:micro}a iii)) shows duodenal glands adjacent to amorphous eosinophilic material. Identical area is imaged on the Congo red stain (Fig.~\ref{fig:micro}a i)) with amorphous eosinophilic material confirmed to be amyloid. We measure the birefringence contrast as described in section~\ref{meth} and shown in Fig.~\ref{fig:s12_bir}a,c for both the Congo red and H\&E stained duodenal tissue section. We analyze high contrast regions R0 and R1 in Fig.~\ref{fig:s12_bir}b,d and observe the intensity oscillations (from Eq.~\ref{T}) as a function of polarizer angle. Clearly, for both the Congo red and H\&E stained sample, the characteristic $k=4$ Fourier frequency component of birefringence structure dominates the signal. Also, as expected, birefringence contrast is visible in the H\&E stained samples with reduced contrast compared to Congo red as Congo red enhances intrinsic birefringence already present in any structures, amyloid fibrils and otherwise~\cite{wolman1965cause, taylor1974determination, howie2009optical, howie2019origins}. We take reference measurements of normal duodenal tissue, as shown in Fig.~\ref{fig:s12_normal},~\ref{fig:s12_nobir} in the Appendix, and we observe no birefringence in the normal tissue sample as expected.

We fit the data with Eq.~\ref{eq_fit} and provide the fit parameters for the duodenal tissue sample in the Appendix in Table.~\ref{stomach_table} for reference. From Eq.~\ref{deltan}, we quantify the birefringence of the sample $\Delta n$ as mentioned in Ref.~\cite{westreich2019novel}. For the Congo red and H\&E stained duodenal tissue sample, $\Delta n$ for the R0 and R1 regions are computed from the fit and summarized in Table.~\ref{bir_table}. We observe a consistent significant difference between the computed $\Delta n$ values for Congo red and H\&E stained samples due to the large expected differences in observed contrast depending on the staining method~\cite{wolman1965cause, taylor1974determination, howie2009optical, howie2019origins}. For example, for R0 of the Congo red (H\&E) stained sample, the refractive index difference between the fast and slow axis is $\Delta n = 0.0093\pm0.0003$ ($\Delta n = 0.0044\pm0.0001$). From these measurements we also quantify relative amounts of amyloid present in the samples as described in section~\ref{meth} and summarized in Table.~\ref{amy_table}. The density of amyloid metric compares variations in brightness as determined by the $B_1$ amplitude in Eq.~\ref{eq_fit}, similar to the methods used in Ref.~\cite{westreich2019novel}. Notably, since Congo red enhances the contrast of birefringent structures, this metric predicts that all Congo red stained samples contain more amyloid then the H\&E stained samples. Thus, we conclude that this metric for quantifying relative amount of amyloid is well-suited for comparing density across different samples with the same staining procedure and should not be used to make claims on amyloid amount between Congo red and H\&E.

The computed birefringence values are further modified due to the linear dichroism term, as seen in Eq.~\ref{eq_fit},~\ref{eq_dic} and described in section~\ref{meth}. From the measurements, we quantify $\Delta k$, the normalized difference between transmitted intensities for the primary polarizations, and summarize the results in Table~\ref{dic_table}. We observe that the dichroism contribution is comparable for both the Congo red and H\&E stained samples, indicating that, as expected, Congo red primarily enhances birefringence.

We analogously characterize the other biological tissue samples and observe similar results whereby the birefringence of amyloid is detectable without using Congo red staining. Shown in Fig.~\ref{fig:micro}b i)-iv) and Fig.~\ref{fig:s9_bir} are microscope images and birefringence contrast for an abdominal subcutaneous tissue sample.  H\&E image (Fig.~\ref{fig:micro}b iii)) shows adipose tissue infiltrated by amorphous eosinophilic material. Identical area is imaged on the Congo red stain (Fig.~\ref{fig:micro}b i)) with amorphous eosinophilic material confirmed to be amyloid. Once again, we observe birefringence in both the Congo red and H\&E stained samples, with lower contrast in the H\&E stained sample as expected. 

Shown in Fig.~\ref{fig:micro}c i)-iv) and Fig.~\ref{fig:s10_bir} are the microscope images and birefringence contrast for the thyroid medullary carcinoma sample. H\&E image (Fig.~\ref{fig:micro}c iii)) shows thyroid medullary carcinoma cells embedded in eosinophilic amorphous eosinophilic material.  Identical area is imaged on the Congo red stain (Fig.~\ref{fig:micro}c i)) with amorphous eosinophilic material confirmed to be amyloid. 
Shown in Fig.~\ref{fig:micro}d i)-iv) and Fig.~\ref{fig:s11_bir} are the microscope images and birefringence contrast for the breast tissue sample.  H\&E image (Fig.~\ref{fig:micro}d iii)) shows breast stromal tissue infiltrated by amorphous eosinophilic material. Identical area is imaged on the Congo red stain (Fig.~\ref{fig:micro}d i)) with amorphous eosinophilic material confirmed to be amyloid. The computed birefringence for all samples and relative amyloid metric are summarized in Tables.~\ref{bir_table},~\ref{amy_table}.

For further analysis of the optical properties of the samples, we fix the polarizer orientation and only rotate the analyzer to compute the optical activity of the sample, as shown in Fig.~\ref{fig:s9_opt} for the abdominal subcutaneous tissue sample. In some cases, such as point R1 of the H\&E stained sample, we observe statistically significant sample induced rotation of several degrees, indicating an optically active sample. Typically, the optical rotation $\chi$ is near $0\degree$ with several degrees of uncertainty, indicating that there is no significant optical activity. The optical activity results are summarized in Table.~\ref{opt_table}.

In Fig.~\ref{fig:freqs} we identify the contrast of the two main Fourier harmonics for this analysis, birefringence and dichroism, for all measurements in Figs.~\ref{fig:s12_bir},~\ref{fig:s9_bir},~\ref{fig:s10_bir},~\ref{fig:s11_bir}. We observe that Congo red stained samples exhibit larger birefringence harmonics than all the H\&E stained samples whereas the dichroism harmonic contrast varies between the two different stains. Furthermore, we compare the Fourier spectrum for the lowest birefringence contrast ROI (H\&E R0 from Fig.~\ref{fig:s9_bir}) with normal tissue (Congo red R1 from Fig.~\ref{fig:s12_normal} in the Appendix) and observe that H\&E R0 maintains a dominating birefringence harmonic in comparison to other frequencies (namely $k= 1,~3$), while the normal tissue spectrum is much noisier.

\begin{figure*}
    \centering\includegraphics[width=1\linewidth]{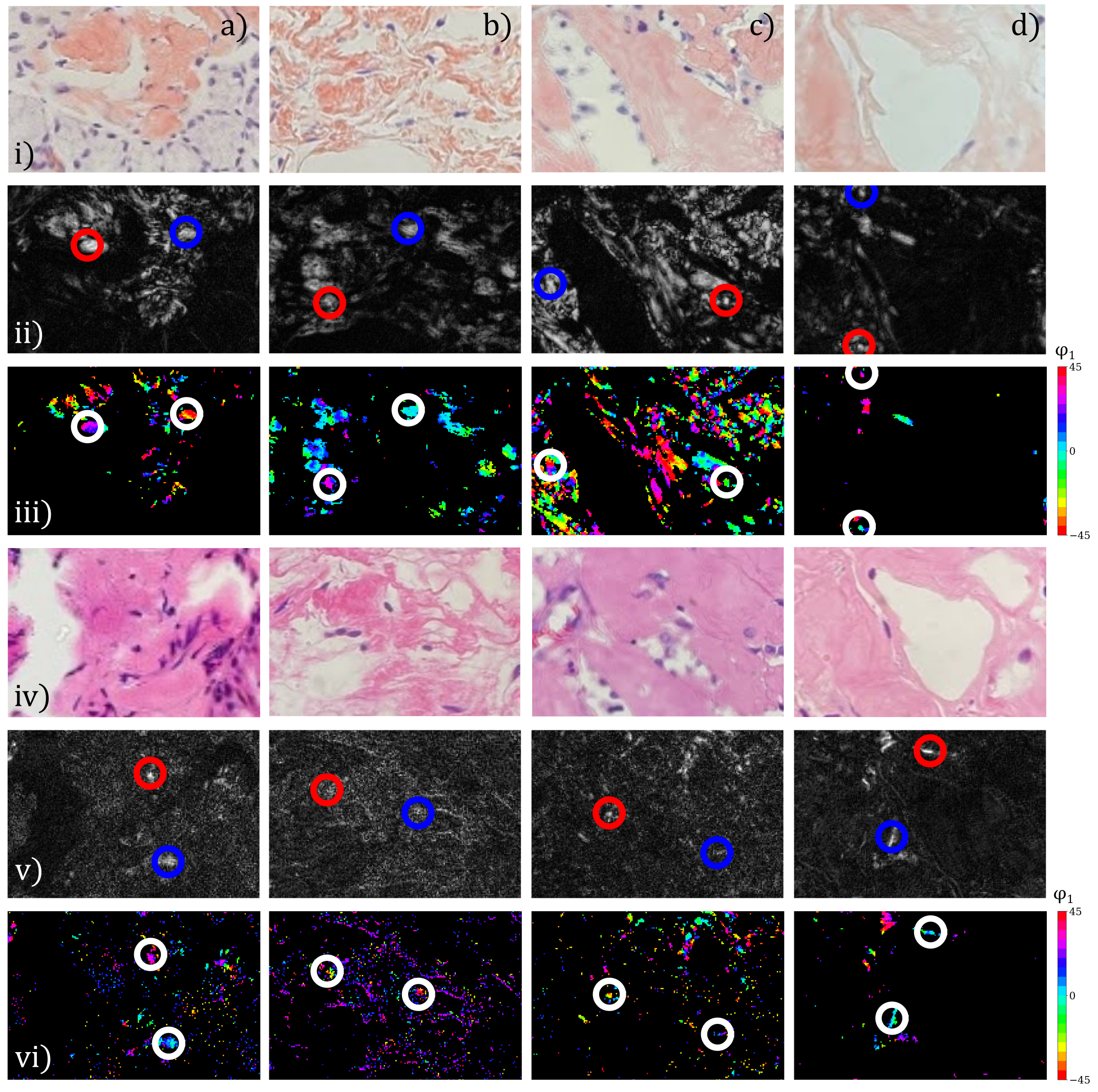}
    \caption{Analysis of birefringence structure orientation for all samples, alongside corresponding microscope images and birefringence contrast maps for comparison depicting identical areas. Columns a-d) are the duodenal, abdominal subcutaneous tissue, thyroid, and breast tissue samples respectively. Rows i-iii) (iv-vi)) are the microscope images, birefringence contrast, and birefringence phase results for Congo red (H\&E) stained samples. The birefringent phase maps display $\phi_1$ from Eq.~\ref{eq_fit} from $-45\degree~to~45\degree$ in $5\degree$~bins in a focused region around the R0, R1 points for all samples. We restrict the phase maps to statistically significant values by comparing to birefringence contrast values above 0.3 for Congo red and 0.1 for H\&E. Notably, the areas with a stronger birefringent structures exhibit regions of similar phase, i.e. similar birefringent axis aliment and thus likely tissue structural alignment.}
    \label{fig:bir_phase}
\end{figure*}

Furthermore, for all samples, we both qualitatively and quantitatively analyze birefringent orientation of the amyloid structure. Shown in Fig.~\ref{fig:bir_phase}, are the birefringent phase maps ($\phi_1$ in Eq.~\ref{eq_fit} from $-45\degree~to~45\degree$) in a focused region around the R0, R1 points for all samples. We observe that regions with birefringent structures exhibit areas of uniform phase, caused by tissue structure alignment. Presented in Table.~\ref{bir_phase_table} are the calculated birefringent axis orientations in the 4x4 pixel bins in the ROIs indicated by white circles in Fig.~\ref{fig:bir_phase}.

\begin{figure*}
    \centering\includegraphics[width=1\linewidth]{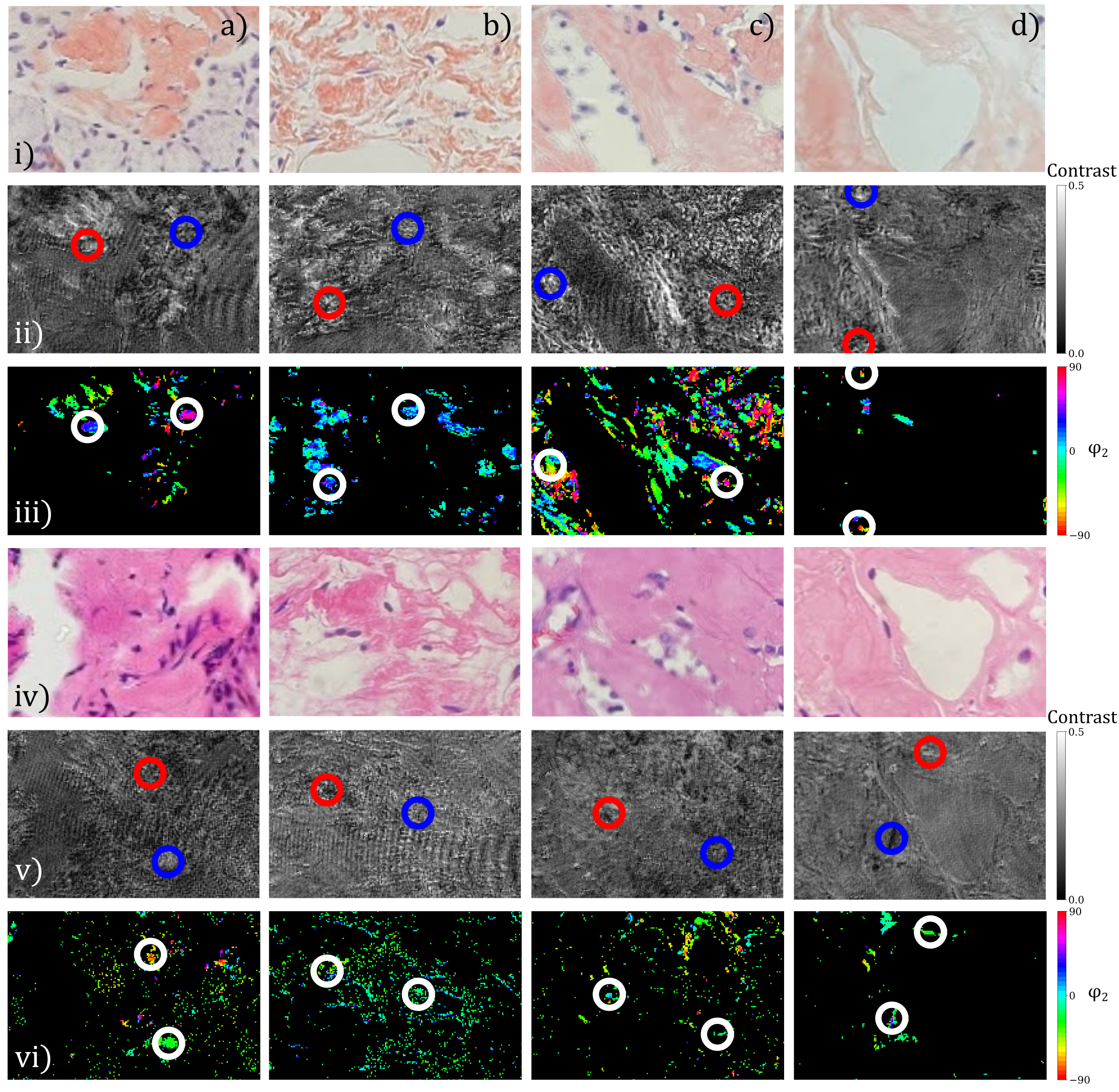}
    \caption{Analysis of dichroism maximum absorption orientation for all samples, alongside corresponding microscope images and dichroism/absorption contrast maps for comparison depicting identical areas. Columns a-d) are the duodenal, abdominal subcutaneous tissue, thyroid, and breast tissue samples respectively. Rows i-iii) (iv-vi)) are the microscope images, absorption contrast, and absorption phase results for Congo red (H\&E) stained samples. The absorption phase maps display $\phi_2$ from Eq.~\ref{eq_fit} from $-90\degree~to~90\degree$ in $5\degree$~bins in a focused region around the R0, R1 points for all samples. We restrict the phase maps identically to Fig.~\ref{fig:bir_phase}.}
    \label{fig:abs_phase}
\end{figure*}

We analogously analyze the dichroism contribution and measure dichroism phase maps representing the orientation of the most absorbing direction as shown in Fig.~\ref{fig:abs_phase}. We plot the absorption phase maps ($\phi_2$ in Eq.~\ref{eq_fit} from $-90\degree~to~90\degree$) in a focused region around the R0, R1 points for all samples and summarize the corresponding dichroism maximum absorption orientation (from Eq.~\ref{eq_dic}) in Table.~\ref{abs_phase_table}. We observe that ROIs with significant dichroism modulation of the dominant $k=4$ birefringent oscillations (for example, see Fig.~\ref{fig:s12_bir} HE R1, Fig.~\ref{fig:s9_bir} HE R0, R1, CR R1, Fig.~\ref{fig:s10_bir} HE R0, R1, Fig.~\ref{fig:s11_bir} CR R1, HE R1) maintain similar $\phi_1,~\phi_2$ values from Eq.~\ref{eq_fit}, typically within uncertainty, confirming the relationship between dichroism and birefringence axis orientations~\cite{tuchin2016polarized, jin2003imaging}, which have been indicated to be originated from the same microstructure and overlapping with each other~\cite{tuchin2016polarized}.

\begin{table*}[]
\centering
\begin{tabular}{ |p{4.5cm}||p{3cm}|p{3cm}|p{3cm}|p{3cm}|p{3cm}|  }
 \hline
 \multicolumn{5}{|c|}{Birefringence ($\Delta n$)} \\
 \hline
 Sample & Congo red, R0 & Congo red, R1 & H\&E, R0 & H\&E, R1 \\
 \hline
 Duodenal tissue               & $0.0093\pm0.0003$ & $0.0076\pm0.0002$ & $0.0044\pm0.0001$ & $0.0033\pm0.0001$ \\
 Abdominal subcutaneous tissue & $0.0066\pm0.0001$ & $0.0078\pm0.0002$ & $0.0052\pm0.0003$ & $0.0042\pm0.0006$ \\
 Thyroid tissue               & $0.0078\pm0.0003$ & $0.0094\pm0.0002$ & $0.0039\pm0.0002$ & $0.0034\pm0.0002$ \\
 Breast tissue                & $0.0065\pm0.0002$ & $0.0113\pm0.0002$ & $0.0028\pm0.0001$ & $0.0032\pm0.0001$ \\
 \hline
\end{tabular}
\caption{Computed birefringence $\Delta n$ values (Eq.~\ref{deltan}) for the duodenal, abdominal subcutaneous, thyroid, and breast tissue samples.}
\label{bir_table}
\end{table*}

\begin{table*}[]
\centering
\begin{tabular}{ |p{4.5cm}||p{3cm}|p{3cm}|p{3cm}|p{3cm}|p{3cm}|  }
 \hline
 \multicolumn{5}{|c|}{Relative amount of amyloid} \\
 \hline
 Sample & Congo red, R0 & Congo red, R1 & H\&E, R0 & H\&E, R1 \\
 \hline
 Duodenal tissue               & $0.69\pm0.08$ & $0.45\pm0.06$ & $0.15\pm0.02$ & $0.09\pm0.01$ \\
 Abdominal subcutaneous tissue & $0.35\pm0.04$ & $0.49\pm0.06$ & $0.22\pm0.03$ & $0.14\pm0.04$ \\
 Thyroid tissue               & $0.48\pm0.06$ & $0.69\pm0.08$ & $0.12\pm0.02$ & $0.09\pm0.01$ \\
 Breast tissue                & $0.33\pm0.04$ & $1\pm0.11$ & $0.06\pm0.01$ & $0.08\pm0.01$ \\
 \hline
\end{tabular}
\caption{Quantifying the relative amount of amyloid between the different samples and different staining methods as described in section~\ref{meth}. The normalized metric ranges from 0 (`no amyloid') to 1 (`high density of amyloid'). Note that the enhanced birefringent contrast with Congo red samples significantly alters the determination of amount of amyloid with this metric and thus it is best to compare the stains separately.}
\label{amy_table}
\end{table*}

\newpage

\begin{table*}[]
\centering
\begin{tabular}{ |p{4.5cm}||p{3cm}|p{3cm}|p{3cm}|p{3cm}|p{3cm}|  }
 \hline
 \multicolumn{5}{|c|}{Linear Dichroism ($\Delta k$)} \\
 \hline
 Sample & Congo red, R0 & Congo red, R1 & H\&E, R0 & H\&E, R1 \\
 \hline
 Duodenal tissue               & $0.00043 \pm 0.00006$ & $0.00019 \pm 0.00004$ & $0.00006 \pm 0.00001$ & $0.00010\pm 0.00001$ \\
 Abdominal subcutaneous tissue & $0.00014 \pm 0.00002$ & $0.00051 \pm 0.00004$ & $-0.00017 \pm 0.00004$ & $-0.00017 \pm  0.00006$ \\
 Thyroid tissue               & $-0.00024 \pm 0.00005$ & $-0.00037 \pm 0.00005$ & $-0.00014 \pm 0.00002$ & $-0.00006 \pm 0.00002$ \\
 Breast tissue                & $0.00019 +\- 0.00003$ & $0.00060 +\- 0.00006$ & $-0.00010 +\- 0.00001$ & $0.000031 +\- 0.000008$ \\
 \hline
\end{tabular}
\caption{Computed linear dichroism $\Delta k$ values (Eq.~\ref{eq_dic}) for the duodenal, abdominal subcutaneous, thyroid, and breast tissue samples.}
\label{dic_table}
\end{table*}

\begin{table*}[]
\centering
\begin{tabular}{ |p{4.5cm}||p{3cm}|p{3cm}|p{3cm}|p{3cm}|  }
 \hline
 \multicolumn{5}{|c|}{Birefringent Axis Orientation} \\
 \hline
 Sample & Congo red, R0 & Congo red, R1 & H\&E, R0 & H\&E, R1 \\
 \hline
 Duodenal tissue               & $(-67\degree~or~23\degree)\pm1\degree$ & $(-38\degree~or~52\degree)\pm1\degree$ & $(-66\degree~or~24\degree)\pm1\degree$  & $(-83\degree~or~7\degree)\pm1\degree$ \\
Abdominal subcutaneous tissue & $(-63\degree~or~27\degree)\pm1\degree$ & $(-3\degree~or~87\degree)\pm1\degree$  &  $(-33\degree~or~57\degree)\pm2\degree$ & $(-64\degree~or~26\degree)\pm4\degree$ \\
 Thyroid tissue               & $(-12\degree~or~78\degree)\pm1\degree$ & $(-53\degree~or~37\degree)\pm1\degree$ & $(-34\degree~or~56\degree)\pm1\degree$  & $(-82\degree~or~8\degree)\pm2\degree$ \\
 Breast tissue                & $(-4\degree~or~86\degree)\pm1\degree$  & $(-65\degree~or~25\degree)\pm1\degree$ & $(-87\degree~or~3\degree)\pm1\degree$   & $(-88\degree~or~2\degree)\pm1\degree$ \\
 \hline
\end{tabular}
\caption{Computed birefringent orientation for R0, R1 regions of the Congo red and H\&E stained tissue samples. From Eq.~\ref{eq_fit}, the first maxima of the birefringent component occurs when the polarizer angle equals $-\phi_1$ which implies that the birefringent orientation is $-\phi_1 \pm 45\degree$ from Eq.~\ref{deltan}. The birefringent axis orientation maps corresponding to these values are in Fig.~\ref{fig:bir_phase} in 5\degree~bins from $-45\degree~to~45\degree$.}
\label{bir_phase_table}
\end{table*}

\begin{table*}[]
\centering
\begin{tabular}{ |p{4.5cm}||p{3cm}|p{3cm}|p{3cm}|p{3cm}|  }
 \hline
 \multicolumn{5}{|c|}{Maximum absorption Orientation ($\phi_2'$)} \\
 \hline
 Sample & Congo red, R0 & Congo red, R1 & H\&E, R0 & H\&E, R1 \\
 \hline
 Duodenal tissue              & $50\degree\pm4\degree$ & $24\degree\pm1\degree$ & $5\degree\pm7\degree$ & $25\degree\pm3\degree$ \\
Abdominal subcutaneous tissue & $45\degree\pm4\degree$ & $-35\degree\pm2\degree$ & $-17\degree\pm6\degree$ & $1\degree\pm10\degree$ \\
 Thyroid tissue               & $-53\degree\pm6\degree$ & $38\degree\pm4\degree$ & $-10\degree\pm3\degree$ & $32\degree\pm8\degree$ \\
 Breast tissue                & $2\degree\pm4\degree$ & $14\degree\pm3\degree$ & $28\degree\pm3\degree$ & $45\degree\pm8\degree$ \\
 \hline
\end{tabular}
\caption{Computed dichroism maximum absorption orientation for R0, R1 regions of the Congo red and H\&E stained tissue samples from Eq.~\ref{eq_fit},~\ref{eq_dic}. The dichroism maximum absorption orientation maps corresponding to these values are in Fig.~\ref{fig:abs_phase} in 5\degree~bins from $-90\degree~to~90\degree$.}
\label{abs_phase_table}
\end{table*}

\begin{table*}[]
\centering
\begin{tabular}{ |p{4.5cm}||p{3cm}|p{3cm}|p{3cm}|p{3cm}|p{3cm}|  }
 \hline
 \multicolumn{5}{|c|}{Optical Activity ($\chi, \Delta n_{OA}$)} \\
 \hline
 Sample & Congo red, R0 & Congo red, R1 & H\&E, R0 & H\&E, R1 \\
 \hline
Abdominal subcutaneous tissue & $-2\degree\pm 2\degree$\newline $-0.0008\pm0.0008$ & $-0.5\degree\pm 1\degree$\newline $-0.0002\pm0.0005$ & $0\degree\pm 2\degree$\newline $0.0000\pm 0.0005$ & $-6\degree\pm 1\degree$\newline $0.0021\pm0.0004$ \\
 Thyroid tissue & $0.5\degree\pm 1\degree$\newline $0.0002\pm0.0005$ & $0.7\degree\pm 2\degree$\newline $0.0002\pm0.0008$ & $3\degree\pm 1\degree$\newline $0.0009\pm 0.0003$ & $-1\degree\pm 3\degree$\newline $-0.0004\pm0.001$ \\
 Breast tissue & $3\degree\pm 2\degree$\newline $0.0012\pm0.0007$ & $-3\degree\pm 1\degree$\newline $-0.0012\pm0.0005$ & $2\degree\pm 1\degree$\newline $0.0007\pm 0.0005$ & $-0.6\degree\pm 0.8\degree$\newline $-0.0002\pm0.0003$ \\
 \hline
\end{tabular}
\caption{Computed optical activity induced phase shift ($\chi$) and $\Delta n_{OA}$ (Eq.~\ref{delta_opt}) values for the abdominal subcutaneous tissue, thyroid, and breast tissue samples.}
\label{opt_table}
\end{table*}

\section{Conclusion and Discussion}
In this work we have implemented a polarized light microscopy system capable of detecting and quantifying birefringent material such as amyloid in biological samples without the need for Congo red staining. We compare birefringence maps of Congo red and H\&E stained duodenal tissue, abdominal subcutaneous tissue, thyroid tissue, and breast tissue samples and clearly observe the $k = 4$ Fourier frequency component characteristic of birefringent structure. We also quantify birefringent axis orientation, optical activity, amyloid fibril density, and linear dichroism for the various samples. The presented techniques provide a solution to the complications regarding Congo red stained samples for the purpose of detecting amyloid birefringence. These quantitative measurements are objective and more robust metrics of amyloid in tissue, compared to Congo Red, and may correlate with severity of disease state, as shown for p53 amyloid in cancer pathology~\cite{sengupta2023p53, iwahashi2022cytoplasmic}.

A possible disadvantage of this work, similar to PLM with Congo red, is that our measurements will detect birefringence and optical activity of any structures within the sample and is not amyloid fibril specific. Therefore, as done in this work, trained pathologists cross reference microscope images with birefringent contrast maps for verification of amyloids within samples. 

This work opens the door for many future studies with polarized light microscopy for amyloid diagnosis. For example, these imaging techniques might be applied for thin unstained pathology slides to detect amyloid fibrils entirely independent of staining methods. Furthermore, this work is well suited to incorporate analysis with structured light~\cite{rubinsztein2016roadmap}. This promising next step extends techniques beyond polarized light to include an additional degree of freedom, orbital angular momentum, that is capable of probing the structure of samples, such as radially arranged amyloid fibrils~\cite{jin2003imaging, kelenyi1967thioflavin}. 

\section*{Acknowledgments}

This work was supported by the Canadian Excellence Research Chairs (CERC) program, the Natural Sciences and Engineering Research Council of Canada (NSERC), and the Canada  First  Research  Excellence  Fund  (CFREF), the Canadian Cancer Society, the Canadian Institutes of Health Research - Institute of Cancer Research, and Brain Canada Foundation through a CCS/CIHR/BC Spark Grant.

\newpage
\clearpage
\bibliography{mybib}

\clearpage
\onecolumngrid
\section*{Appendix}

\begin{figure*}[h]
    \centering\includegraphics[width=1\linewidth]{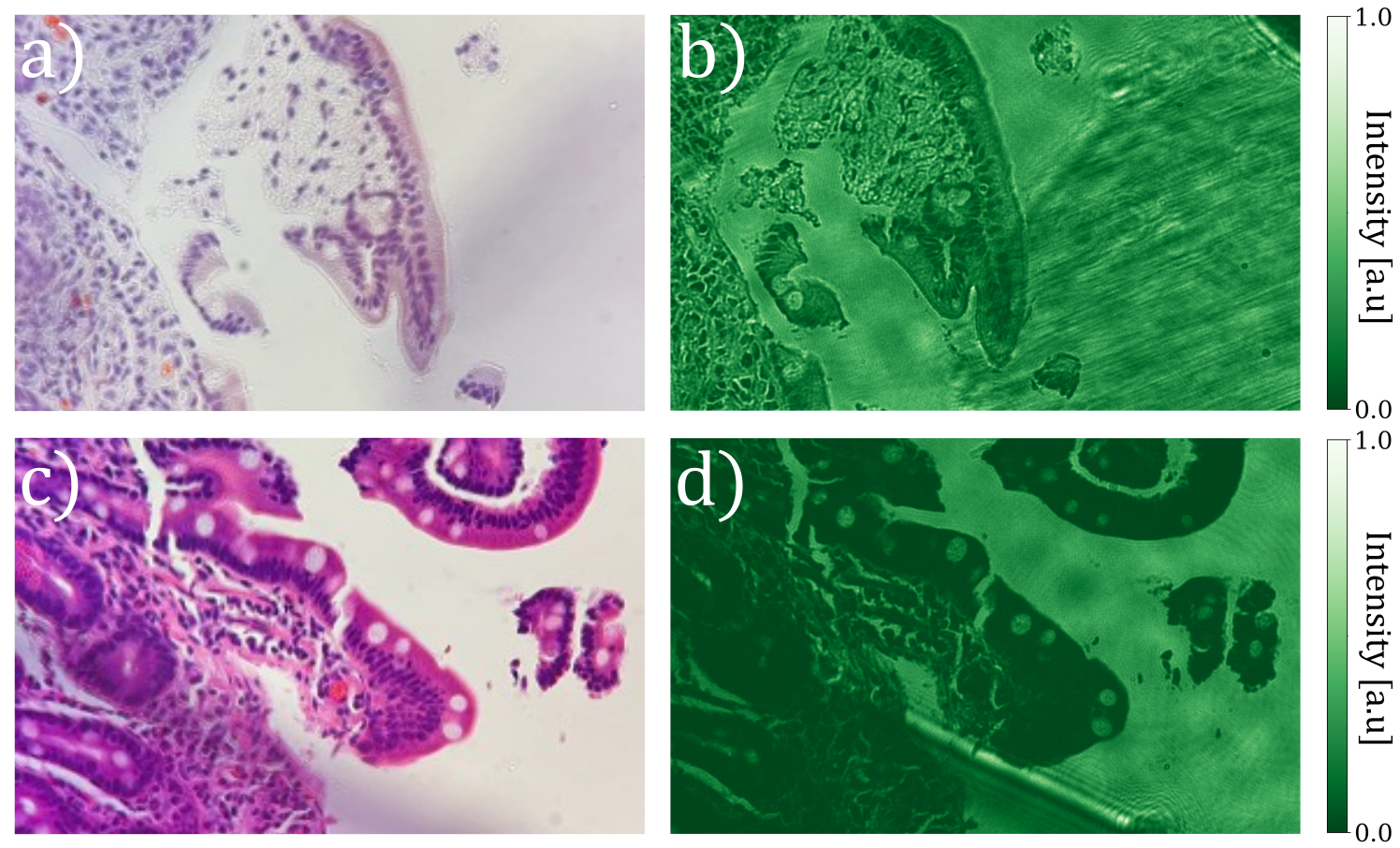}
    \caption{a) Microscope image at 40x magnification for a Congo red stained normal duodenal sample. b) The corresponding image taken with our optical system for parallel polarizers. c-d) Analogous images as a-b) for H\&E stained normal duodenal tissue.}
    \label{fig:s12_normal}
\end{figure*}

\begin{figure*}
    \centering\includegraphics[width=1\linewidth]{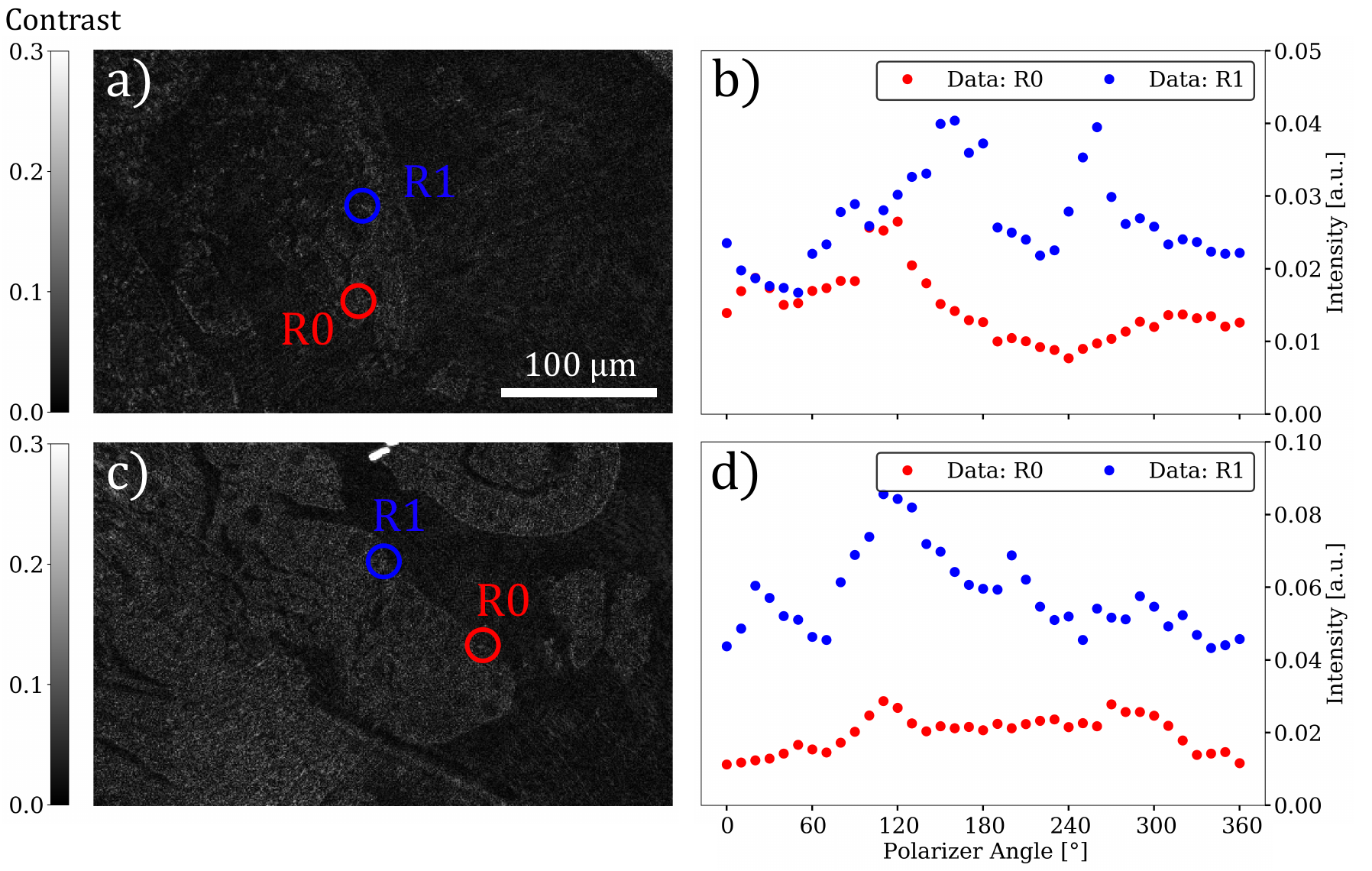}
    \caption{a) The computed birefringence contrast for the Congo red stained thyroid tissue sample. Labels R0 and R1 correspond to analyzed pixels in b), plotting measured intensity T from Eq.~\ref{T} as a function of polarizer angle. No dominant Fourier frequency components are observed since this region is simply normal tissue with no birefringence. In c-d) are the analogous figures to a-b) for and H\&E stained sample. Results are agreeable with the Congo red stained sample.}
    \label{fig:s12_nobir}
\end{figure*}

Pathologists identify regions of the same tissue sample that are normal tissue and show no amyloid. These regions are used as reference measurements to compare with regions of interest pathologists determine to contain amyloid. Shown in Fig.~\ref{fig:s12_normal}a-d) are images of normal duodenal tissue stained with Congo red or H\&E under a microscope, and the corresponding images taken with our custom setup when the polarizers are parallel. We run the measurement algorithm to acquire all experimental images and compute the birefringence contrast as shown in Fig.~\ref{fig:s12_nobir}a-d). As expected, no birefringent structure is observed in either the Congo red or H\&E stained duodenal samples.

Fit parameters from Eq.~\ref{eq_fit} for the duodenal tissue sample are provided in Table.~\ref{stomach_table} for reference.

\begin{table*}[h]
\centering
\begin{tabular}{ |p{2.5cm}||p{2.5cm}|p{2.5cm}|p{2.5cm}|p{2.5cm}|p{2.5cm}|  }
 \hline
 \multicolumn{6}{|c|}{Stomach Sample Fit Parameters} \\
 \hline
 Stain, R\# & A & $B_1$ & $\phi_1$ & $B_2$ & $\phi_2$ \\
 \hline
 Congo red, R0     & $0.075\pm 0.002$ & $0.050\pm 0.003$ & $-22\degree\pm 1\degree$ & $0.020\pm 0.003$ & $-50\degree\pm 4\degree$\\
 Congo red, R1     & $0.050\pm 0.002$ & $0.033\pm 0.002$ & $7\degree\pm 1\degree$ & $0.009\pm 0.002$ & $-24\degree\pm 6\degree$\\
 H\&E, R0 & $0.0250\pm 0.0005$ & $0.0111\pm 0.0007$ & $-21\degree\pm 1\degree$ & $0.0029\pm 0.0007$ & $-4\degree\pm 6\degree$ \\
 H\&E, R1 & $0.0134\pm 0.0004$ & $0.0062\pm 0.0005$ & $-38\degree\pm 1\degree$ & $-0.0048\pm 0.0005$ & $-25\degree\pm 3\degree$\\
 \hline
\end{tabular}
\caption{Fit parameters from Eq.~\ref{eq_fit} for the duodenal tissue sample used for computing birefringence, birefringence orientation, relative amyloid density, and dichroism.}
\label{stomach_table}
\end{table*}

\end{document}